\documentclass[
 reprint,
 amsmath,amssymb,superscriptaddress,
 aps,prl
]{revtex4-1}
\usepackage{hyperref}

\usepackage{graphicx}
\usepackage{dcolumn}
\usepackage{bm}
\usepackage{dsfont}
\usepackage{natbib}
\usepackage[usenames, dvipsnames]{color}

\newcommand{\ket}[1]{|#1\rangle}
\newcommand{\bra}[1]{\langle #1|}

\hypersetup{colorlinks=true, citecolor=blue}

\begin{document}

\title{Classical and Nonclassical Time Dilation for Quantum Clocks}

\author{A.~J.~Paige}\thanks{a.paige16@imperial.ac.uk}
\author{A.~D.~K. Plato}
\author{M.~S. Kim}
\affiliation{QOLS, Blackett Laboratory,\\ Imperial College London,\\ South Kensington, London, SW7 2AZ, UK.}

\begin{abstract}
Proper time, ideal clocks, and boosts are well understood classically, but subtleties arise in quantum physics. We show that quantum clocks set in motion via momentum boosts do not witness classical time dilation. However, using velocity boosts we find the ideal behaviour in both cases where the quantum clock and classical observer are set in motion. Without internal state dependent forces additional effects arise. As such, we derive observed frequency shifts in ion trap atomic clocks, indicating a small additional shift, and also show the emergence of non-ideal behaviour in a theoretical clock model.
\end{abstract}
\maketitle

Ideal clocks and proper time are key concepts in
special and general relativity \cite{Rindler2001}. Full understanding of the union between relativity and quantum mechanics, must include how these ideas extend to the quantum realm. Recent work in this area can broadly be divided by whether the quantum clocks follow classical or quantum trajectories.

Adopting the former approach \cite{Lock2016,Lorek2015,Lindkvist,Lindkvist2015} enables the utilization of techniques from quantum field theory in curved spacetime. In particular this has allowed explorations into consequences of the Unruh effect \cite{Unruh1976}, and applications of techniques from relativistic quantum metrology \cite{Ahmadi2014,Ahmadi2014a}. On the other hand, for quantum clocks following quantum mechanical trajectories \cite{Zych2011,Pikovski2015,Zych2016,Zych2017,Diosi2015,Adler2015,Bonder2015,Toros2017,Pikovski2017,Bushev2016,Franson2016}, most progress has been made investigating connections between proper time and mass superpositions \cite{Greenberger2001}. This has necessitated the rejection of the Bargmann mass superselection rule \cite{Bargmann1954}, on the grounds that our universe is not Galilean. Notably this paradigm was used to investigate ideas for intrinsic time dilation decoherence caused by gravity \cite{Pikovski2015}.

In this work we follow the second approach, where the clock's motion is described quantum mechanically. We show that a quantum clock set into motion by a force that does not depend on the internal state is not witnessing classical time dilation. This is because quantum clocks require coherence in some non-degenerate energy states \cite{Castro-Ruiz2015,Kwon2017} but the inertial mass of this energy means that assigning an identical momentum to each branch of the superposition does not correspond to a well defined velocity. We therefore show that momentum boosts lead to a nonclassical dilation due to the lack of a unique Lorentz factor. On the other hand, by suitably coupling the motional and internal degrees of freedom, one can apply a velocity boost which exactly recovers the expected classical time dilation results for the ``twin paradox,'' in both cases where the observer and the quantum clock are respectively the ones set in motion.

We start from a Hamiltonian modified to account for the inertial mass of internal energy. We then consider sequences of appropriately centred boosts and evolution operators to derive the different possible clock behaviours in a twin paradox scenario. From the classical observer's frame we show that the difference between the classical observer and quantum clock being set in motion is captured by translation operators, and that it is the transformation under translation operators that enables the velocity boost to describe both situations. We demonstrate how these translation operators can be understood via considering the placement of the origin for the required potentials. In addition, we show that considering the velocity kick as the classical observer changing frames immediately tells us that entanglement is frame dependent and demonstrate consistency with the equivalence principle. We highlight that without an internal state dependent force one should expect additional effects. We demonstrate this for frequency shifts in ion traps, predicting the already observed shifts and an additional smaller shift. We also analyse the effect on a Salecker-Wigner-Peres clock \cite{Salecker1958,Peres1980} finding non-ideal clock behaviour.

\emph{Modified Hamiltonian ---} We begin by presenting the simplest argument for the modified Hamiltonian. The modification was introduced and used to study quantum mechanical proper time \cite{Zych2011}, and has also been shown to resolve paradoxes in quantum optics \cite{Sonnleitner2017,Barnett2018}. For a free composite particle of mass $M,$ the non-relativistic Hamiltonian will consist of a kinetic energy term $\frac{p^2}{2M},$ and an internal energy term $H_0.$  However the internal energy should contribute to the inertial mass since special relativity dictates that energy and inertial mass are equivalent. This leads to $M=m+H_{0}/c^{2},$ where we take $m$ as the rest mass of the particle in its internal energy ground state $|0\rangle,$ and set $H_0|0\rangle=0.$ Thus we have
\begin{equation} \label{eq:FullHamiltonian}
    \begin{split}
    H & =  \frac{{p}^{2}}{2M}+{H}_{0},  \\ 
   & = \frac{p^2}{2m} + H_{0}\left(1-\frac{p^2}{2mMc^2}\right).
\end{split}
\end{equation}
To arrive at the second line we have used $1/(x+y) = 1/x - y/x(x+y).$ Note that since $M$ is now operator valued, there is potential for ambiguity with operator ordering. However if the internal Hamiltonian commutes with the total momentum $[H_0,p]=0,$ then $[M^{-1},p]=0.$ Fully accounting for relativity would strictly imply that the internal degrees of freedom should be described by a relativistic wave equation (or a quantum field theory). However the approach here is that regardless of the formalism, the effect on the centre of mass dynamics should only be via a mass change, otherwise we could not claim that energy and inertial mass are equivalent.

This Hamiltonian is often expanded in ${H}_{0}/mc^{2}$ neglecting higher order terms \cite{Pikovski2015,Barnett2018}. It is then tempting to claim that $(1-\frac{p^2}{2m^2c^2})$ represents our familiar notion of time dilation, however, this is not correct as will be shown later. It is also important to emphasise that we shall always be working in the limit where the energy $E=\sqrt{p^2c^2 +M^2c^4}$ is approximated by $E=p^2/2M + Mc^2.$ In words, one can consider the regime we utilise as one where the mechanics can be treated in a ``Newtonian'' sense, but the rest mass of the internal energy is now accounted for. On a technical level one must appreciate (as noted in \cite{Diosi2015}) that there are two relevant small quantities: $H_0/mc^2$ and $p^2/m^2c^2,$ where the first relates to the internal degrees of freedom and the second can be viewed as a motional $v^2/c^2$ term (note $p^2/m^2c^2>p^2/M^2c^2,$). It is therefore not sufficient to merely think of approximations in terms of how many factors of $1/c^2,$ are present. One can consider the regime where $H_0/mc^2 \ll 1,$ but $H_0/mc^2 \gg p^2/m^2c^2.$  With this in mind one arrives at Eq. (\ref{eq:FullHamiltonian}) (plus ground state rest mass energy) by expanding the full relativistic energy $E=\sqrt{M^2c^4+p^2c^2}= \gamma Mc^2,$ neglecting terms of $O(p^4/m^4c^4)$ in the Lorentz factor whilst retaining the $H_0/mc^2$ terms and translating the zero of energy by $mc^{2}.$ Note that in practise the order of $H_0/mc^2$ terms kept would be dictated by the physical system under consideration and depends on the integer $n$ for which $(H_0/mc^2)^n \gg p^2/m^2c^2.$ However, as will become apparent, it will prove more straightforward for our initial theoretical study to work with the untruncated $(1+H_0/mc^2)^{-1}.$ We shall denote the unitary evolution generated by Eq. (\ref{eq:FullHamiltonian}) over time $t$ as $U(t).$

\emph{Different boosts ---}
We now explore the consequences of the modified Hamiltonian. Time is a complex topic in quantum mechanics \cite{Busch,Leon2017}, but here we shall simply take the internal state (Hamiltonian $H_0$) to define some quantum clock, and consider the situation where it is boosted away and back, then measured to observe the motion's effect on the clock. To do this we use the following sequence of operations: first we apply some boost operator to the particle and let it freely evolve for some time $t,$ then at a shifted position we apply the inverse boost twice and let it evolve for another time $t,$ and finally we apply the original boost. Note the magnitude of all boosts must be chosen such that the state is kept within the approximation regime of our Hamiltonian. We initially work with the standard quantum mechanical momentum boost, centred at the origin, written as
\begin{equation}
B_{p}(p_{b}) \equiv  e^{ip_{b}x/\hbar}. \label{eq:MomentumBoost}
\end{equation}
This acts on momentum eigenstates as $B_{p}(p_{b})|p\rangle = |p+p_{b}\rangle.$ This represents the physical situation typically considered for use in the laboratory~\cite{loriani2019interference}, with no internal state dependence as per the potentials typically used to move quantum systems (e.g. an ion moved via an Electromagnetic potential).

Using this boost, the translation operator $T(p_{b}t/m)=e^{-ip p_{b}t/m\hbar}$, and the free evolution under the Hamiltonian of Eq. (\ref{eq:FullHamiltonian}), we have
\begin{multline}\label{eq:MomBoostSeq}
    B_{p}(p_{b})U(t)T(p_{b}t/m)B_{p}(-2p_{b})T(-p_{b}t/m)U(t)B_{p}(p_{b}) \\ = e^{-\frac{i2t}{\hbar}\frac{p^2}{2M}}e^{\frac{2it}{\hbar}\frac{p_{b}^2}{2m}}e^{-\frac{2it}{\hbar}H_{0}(1-\frac{p_{b}^2}{2mMc^2})}.
\end{multline}
The first exponential term is the unaltered motional evolution of the state that we would expect if we had not applied any of the boosts. This term has no $p_b$ dependent relativistic corrections, which is not surprising since the adopted formalism does not include the relevant relativistic motional terms. The second term is a global phase that is connected to the choice of the translation operators, discussed in detail below. However, it is the final term that captures the effect on the evolution of the internal state and therefore is our primary focus.

The third exponential term in Eq. (\ref{eq:MomBoostSeq}) has the internal Hamiltonian multiplied by a factor $(1-\frac{p_{b}^2}{2mMc^2})$ that is less than unity. This is precisely the factor we would get by considering Eq. \ref{eq:FullHamiltonian} acting on a momentum eigenstate $|p_{b}\rangle.$ We see that if we have some coherence in the internal state and are using it as a clock, then it appears that the clock is running slower. Note it is not dilated by a constant inverse Lorentz factor as in classical relativity. We can see why this is so by asking what Lorentz factor we would expect. The clock has been given a momentum $p_{b}$ but because our clock is in a superposition of energies $E_{n}$, we also have a superposition of different masses $M_{n}=m+E_{n}/c^{2}.$ Hence we can write various velocities and thus Lorentz factors. For example, the $N$ single shot values $\gamma_n=\sqrt{1+p_{b}^2/M_{n}^{2} c^2},$ or the expectation value of an operator $\langle\gamma\rangle.$ Furthermore, Eq. (\ref{eq:MomBoostSeq}) indicates that none of these is correct. Instead the dilation of the phase factor between any two branches $n,m$ is $1-p_f^2/2M_{n}M_{m}c^2,$ so is always bounded by the single shot inverse Lorentz factors for the individual branches. We in particular note that this is \emph{not equivalent} to what one would expect from the analogous classical mixture of Lorentz factors (see Ref.~\cite{Supplemental} for details).

Once we appreciate these problems we can also see that there is ambiguity in the translation operations. It is natural to center the boost back at a distance from the origin that is equal to the relative velocity imparted multiplied by the time it has freely evolved, but if there are multiple velocities then there are multiple such distances. The clock will have been moved by $vt=(p_b/M)t,$ (this can be seen in the boost sequence $B_p(-p_b)U(t)B_p(p_b)$ which generates the position shift operator $e^{-\frac{it}{\hbar} \frac{p p_b}{M}},$ together with the exponential of a kinetic term). Therefore each internal energy defines the shift $p_{b}t/(m+E_n/c^2),$ so we could justifiably choose to use $T(p_{b}t/(m+E_n/c^2)$ for any of the occupied $n.$ This would make the second exponential term in Eq. (\ref{eq:MomBoostSeq}) become $e^{-\frac{it}{\hbar}(\frac{p_{b}^2}{2m}-\frac{p_{b}^2}{(m+E_{n}/c^2)})},$ but as this is just altering a global phase it does not affect the clock.

Finally we point out that even working to first order in $H_0/mc^2$ we get the displacement operator $e^{-\frac{it}{\hbar}\frac{\hat{p}p_b}{m}(1-\frac{H_0}{mc^2})},$  which is still dependent on the internal state. This is because, unlike the Lorentz factors, the velocities imparted on the different masses are still disparate at this level of approximation. Given sufficient time, and some reasonable localization, the different branches of the clock could in principle become completely spatially separate, which is clearly not in keeping with an interpretation that to this level of approximation we can view this as a clock moving along a single trajectory. Note that when we move it away and back (as per usual in a twin paradox scenario), then we can cancel the shift effects and therefore not notice, but that does not remove the clear issue with the single trajectory interpretation. This shows that we cannot avoid the conceptual problems by simply working to lower order in $H_0/mc^2.$

In order to solve the above problems, we instead use the modified boost operator
\begin{equation}
B_{v}(v_{b}) \equiv e^{i(m+H_0/c^2)v_{b}x/\hbar}. \label{eq:VelocityBoost}
\end{equation}
The motivation for this is clear, we are trying to define a unique velocity and thereby a unique Lorentz factor. One can derive its form in the relevant non-relativistic limit (see Ref.~\cite{Supplemental} for details), and also relate it to the extended Galilean boost $G(v,t)=e^{iv(Mx-tp)/\hbar}$ where $M$ is operator valued, via $G(v,t)=U(t)B_{v}(v)U^\dagger(t).$ The position shift for the clock is now uniquely defined to be $v_{b}t,$ and using this we write
\begin{multline} \label{eq:velocityBoostSeq1}
    B_{v}(v_{b})U(t)T(v_{b}t)B_{v}(-2v_{b})T(-v_{b}t)U(t)B_{v}(v_{b}) \\ =  e^{-\frac{2it}{\hbar}\frac{p^2}{2M}}e^{\frac{2it}{\hbar}\frac{mv_{b}^2}{2}}e^{-\frac{2it}{\hbar}H_{0}(1-\frac{v_{b}^2}{2c^2})}.
\end{multline}
As before the first term is the unaltered motional evolution and the second term is a global phase. It is the third term that interests us. From this we see that the clock has run slower by the inverse of the classical Lorentz factor $\gamma^{-1}\approx(1-\frac{v_{b}^2}{2 c^2}).$ This is exactly as required for classical time dilation.

We can go further and consider the situation where the classical observer is the one that is set into motion. This means that the boosts must all be centred at the origin which gives us 
\begin{multline} \label{eq:velocityBoostSeq2}
    B_{v}(-v_{b})U(t)B_{v}(2v_{b})U(t)B_{v}(-v_{b}) \\ =  e^{-\frac{2it}{\hbar}\frac{p^2}{2M}}e^{-\frac{2it}{\hbar}\frac{mv_{b}^2}{2}}e^{-\frac{2it}{\hbar}H_{0}(1+\frac{v_{b}^2}{2c^2})}.
\end{multline}
Here the quantum clock is running faster by the classical Lorentz factor $\gamma\approx(1+\frac{v_{b}^2}{2 c^2}).$ This is again as expected as the classical observer is moving so their clock runs slower. It is satisfying and encouraging that the modified Hamiltonian produces the correct solution to the twin paradox when we use the velocity boost. The key difference in the two cases is caused by the manner in which the velocity boost transforms under translations $T^{-1}(s)B_{v}(v_{b})T(s) = e^{i(m+H_0/c^2)v_{b}s/\hbar}B_{v}(v_{b}).$ Note that we cannot have the same interpretation with momentum boosts due to the fact that a classical observer cannot move in a superposition of velocities.

It is worth a further comment here on the motional term $e^{-\frac{2it}{\hbar}\frac{p^2}{2M}}.$ As stated above this has been left unaffected by the boosting which is due to the adopted approximations. The full relativistic algebra indicates that a correction term $e^{\pm\frac{2it}{\hbar}\frac{p^2 v_b^2}{4Mc^2}}$ is missing. To obtain a fully consistent regime for these equations one requires this term to approximate the identity, thus restricting the wavepacket momentum spread and time $t.$

An immediate consequence of using the velocity boost is the fact that frame changes alter the entanglement between the motional and internal degrees of freedom. Consider $B_v(v_{b})|p\rangle\frac{1}{\sqrt{2}}(|0\rangle+|1\rangle)=\frac{1}{\sqrt{2}}(|p+M_{0}v_{b}\rangle|0\rangle+|p+M_{1}v_{b}\rangle|1\rangle).$  The internal and motional states are separable and maximally entangled for the non-boosted and boosted frames respectively. Similar behaviour has been demonstrated for internal spin degrees of freedom \cite{Czachor1997,Alsing2002,Peres2002,Gingrich2002}, where the entanglement entropy can change under Lorentz transformations \cite{Peres2002}, by virtue of Wigner rotations. Additionally, by considering the classical observer moving one can show consistency with the equivalence principle. We take fixed acceleration $a,$ for time $t$ broken into $n$ time steps $\delta t = \frac{t}{n}.$ For a single time step $\delta t,$ we apply a boost $B_{v}(a\delta t),$ and the evolution $U(\delta t).$ This gives the unitary $(U(\delta t)B_{v}(a\delta t))^{n},$ for which we take $n\to\infty,$ and reverse the Trotter expansion \cite{Suzuki1976} to arrive at a new unitary which defines the Hamiltonian in the accelerating frame as $H = \frac{p^2}{2M}+H_0+aMx.$
This agrees with the result from an alternative derivation for the accelerating frame's Hamiltonian \cite{ZychThesis}, and importantly is the same form as Hamiltonians to describe the composite particle in a gravitational potential \cite{Zych2011,Pikovski2015,Zych2016,Zych2017}. As a final point we note the case of the observer moving was examined in a different manner by Greenberger \cite{Greenberger2001} (see Ref.~\cite{Supplemental} for details).

\emph{Hamiltonian description of translations--}
The translation operators play a key role in the above. To better understand them we can consider the Hamiltonians necessary to enact the boosts on the state. We start with the velocity case.

Consider classical observers Alice and Bob at rest in each other's frames, separated by a distance $v_{b}t.$ Alice initially holds a quantum clock and she sends it to Bob by applying the potential $-\alpha(m+H_0/c^2)x$ for a short time $\Delta t$ such that the full Hamiltonian in this time is
\begin{equation}
    H = \frac{p^2}{2M} + H_0 - \alpha(m+H_0/c^2)x.
\end{equation}
We choose $\alpha$ large and $\Delta t$ small with $\alpha\Delta t = v_{b},$ such that the first two terms are irrelevant and we effectively generate $U(\Delta t)=e^{i(m+H_0/c^2)v_{b}x/\hbar}.$

After time $t$ evolving under the free Hamiltonian, the clock reaches Bob who applies a potential to send it back. Viewed from Alice's frame this potential is $+2\alpha(m+H_0/c^2)(x-v_bt).$ So the full Hamiltonian is 
\begin{equation}
    H = \frac{p^2}{2M} + H_0 + 2\alpha(m+H_0/c^2)(x-v_{b}t).
\end{equation}
This means that the operator generated in the appropriate limit is
\begin{equation}
    e^{-2i(m+H_0/c^2)(x-v_{b}t)/\hbar} = T(v_{b}t)B_{v}(-2v_{b})T(-v_{b}t).
\end{equation}

One can do the same thing for the momentum boosts but there is now an extra subtlety. Namely, that we do not have a uniquely defined position to centre Bob's potential from, but as we can see from $T(L)B_{p}(-2p_{b})T(-L) = e^{-2ip_{b}(x-L)},$ this only alters a global phase. However, this is only true if we insist on a uniquely defined position shift. It is at least formally interesting to note that if we allow for the positioning of Bob's boost back to be dependent on the internal state, such that the translation operator is $T(p_{b}t/M),$ then we find the internal state evolution multiplied by $(1+\frac{p_{b}^2}{2mMc^2}).$ So the clock runs faster, analogously to the boosted classical observer case, but again not by a relevant classical Lorentz factor. It may be that this approach has some interpretation in the emerging topic of quantum reference frames \cite{Giacomini2017}.

\emph{The nonclassical behaviour ---}
We have shown the velocity boost is the relevant operator when dealing with questions of classical proper time for quantum clocks. However, setting a clock in motion in this manner requires an entangling force that couples the internal and motional degrees of freedom, but for physical situations this is often not the case. Under these circumstances the momentum boost behaviour is more relevant. There has been an experimental proposal \cite{Bushev2016} to use a trapped single electron to test for interference effects caused by the Hamiltonian of Eq. (\ref{eq:FullHamiltonian}). Here we take a different direction, by considering trapped ion optical atomic clock frequency shifts, and arguing they already provide corroboration for the modified Hamiltonian and potentially could provide more.

First we outline the basic operation (see \cite{Ludlow2015} for a review). An ion is trapped in a harmonic potential with trap frequency $\omega_{m},$ and the clock reference frequency is obtained by tuning a laser to an electronic transition frequency $\omega_{0}$ of the ion. The laser frequency is varied to maximise the probability of exciting a transition, which standard quantum mechanics predicts will occur when $\omega_{l}\approx\omega_{0}$. However, with relativity the ion's motion will lead to a dilation effect, which manifests in a frequency shift of the transition. The common approach for incorporating this is to apply the classical time dilation formula, substituting the expectation value of the momentum squared to give $\omega_{l}\approx\omega_{0}(1-\frac{\langle\hat{p}^{2}\rangle}{2m^{2}c^{2}}).$ This is found to be in line with experiment \cite{Chen}.

The approach works well, however it is essentially a semi-classical analysis, because we are making no relativistic correction to the quantum mechanical description. A more natural method is to start from the Hamiltonian of Eq. (\ref{eq:FullHamiltonian}). The interaction of an ion with a monochromatic classical laser field is a well documented problem \cite{Leibfried2003}, and adapting the standard approach we derive a differential equation to describe the time evolution (see Ref.~\cite{Supplemental} for details). Under simplifying approximations we find that the frequency shift for an ion initially in the $n$th Fock state is
\begin{equation}\label{eq:frequencyShift}
\omega_{l} \approx \omega_{0}(1-\frac{\hbar\omega_{m}(n+\frac{1}{2})}{2mc^{2}}+\frac{\hbar\omega_{0}\hbar\omega_{m}(n+\frac{1}{2})}{2(mc^{2})^{2}}).
\end{equation}
The first correction term is the same type of shift studied in \cite{Bushev2016} and is broadly in agreement with the semi-classical argument and the observations \cite{Chen}. This provides empirical evidence for the modified Hamiltonian, since it gives a quantum mechanical description for a real world experiment up to the level of precision achieved. The second correction term captures the additional behaviour that we now expect, however it should be taken as illustrative rather than a concrete experimental prediction. Here we have not considered other effects that could be relevant at this precision, such as the higher order $p^2/M^2c^2$ term. To predict such new shifts one should perform full simulations, with all relevant physics, and using experimental parameters. However, we can make estimates based on the terms above and for a typical experiment the new shift would be a factor of $\frac{\hbar\omega_0}{mc^2} \sim 10^{-10}$ smaller than that observed. Thus state of the art experiments are far from observing these effects. While this is discouraging, the key point is that the modified Hamiltonian predicts effects that could lead to new observable consequences.

The nonclassical dilation can be physically relevant, so it is interesting to consider its effect on proposed theoretical clock models. We do this for the Salecker-Wigner-Peres (SWP) definition of a quantum clock \cite{Salecker1958,Peres1980}. Taking the internal energy Hilbert space to be spanned by $N$ non-degenerate energy eigenstates $|n\rangle,$ $n=0,1,...,N-1,$ with equally spaced eigenvalues such that $H_0=\sum_{n}n\hbar\omega_{0}|n\rangle\langle n|.$ The SWP clock is then defined by the $N$ orthogonal states $\ket{w_k} = \frac{1}{\sqrt{N}} \sum_{n=0}^{N-1}e^{-2\pi i kn/N} \ket{n}.$
Initialised in $|w_0\rangle,$ the clock will pass through successive states $|w_k\rangle$ at external times $t_k=k\tau,$ where $ \tau= \frac{2\pi}{N\omega_0}.$ One then defines a clock operator $T_c= \tau \sum_k k \ket{w_k}\bra{w_k},$ with variance $(\Delta T_{c})^2=0$ at times $t_k,$ and $(\Delta T_c)^2\neq0$ in-between.

For this setup we see that the ideal clock behaviour is broken by the non-linear $H_0$ dependence in the nonclassical dilation. There are no longer well defined ticks with $(\Delta T_c)^2=0.$ One can consider defining effective ticks as points of minimum $(\Delta T_c)^2.$ We numerically find that the time between these new ticks in the large $N$ limit can be approximated analytically utilising time-energy uncertainty relations (details in Ref.~\cite{Supplemental}).

\emph{Conclusions ---}
We have shown that there are conceptual problems with viewing momentum boosts as leading to quantum clocks witnessing a classical time dilation. We found that the velocity boost recovers the expected classical behaviour and demonstrated the importance of translation operators in distinguishing the cases of the clock or the observer being set in motion. We showed how this can be understood by considering the Hamiltonians necessary to realise the boosts on the quantum clock. The velocity boost enables simple demonstrations of the frame dependence of entanglement between internal and motional degrees of freedom, and consistency with the gravitational equivalence principle. We emphasised that moving the quantum clock without an internal state dependent force should present additional effects. From a practical point of view we illustrated this with ion trap atomic clocks, finding that the formalism predicts the already observed relativistic frequency shift and indicates an additional small correction. We also considered the effects of the nonclassical dilation for the SWP clock, finding it removes the ideal clock behaviour.

\begin{acknowledgements}
\emph{Acknowledgments ---}
We acknowledge insightful discussions with Geoffrey Penington, Kiran Khosla, and Renato Renner, along with useful comments on the manuscript from Alexander R. H. Smith. This work was funded by the EPSRC Centre for Doctoral Training in Controlled Quantum Dynamics and a Leverhulme Trust Research Grant (Project RPG-2014-055). MSK acknowledges the Royal Society.
\end{acknowledgements}

\end{document}